# Charge Transport in Mixed Metal Halide Perovskite Semiconductors

Satyaprasad P Senanayak[1#], Krishanu Dey[2#], Ravichandran Shivanna[2], Weiwei Li[3,4], Dibyajyoti Ghosh[5], Bart Roose[2], Youcheng Zhang[2,6], Zahra Andaji-Garmaroudi[2], Nikhil Tiwale[7], Judith L. MacManus Driscoll[3], Richard Friend[2], Samuel D Stranks[*2,8], Henning Sirringhaus[*2]

[1]Nanoelectronics and Device Physics Lab, National Institute of Science Education and Research, School of Physical Sciences, OCC of HBNI, Jatni, 752050 India

[2]Optoelectronics Group, Department of Physics, Cavendish Laboratory, JJ Thomson Avenue, Cambridge, CB3 0HE United Kingdom

[3]Department of Materials Science and Metallurgy, University of Cambridge, 27 Charles Babbage Road, Cambridge CB3 0FS, United Kingdom

[4]MIIT Key Laboratory of Aerospace Information Materials and Physics, Nanjing University of Aeronautics and Astronautics, Nanjing 211106, China

[5]Department of Materials Science and Engineering, Department of Chemistry, Indian Institute of Technology Delhi, Hauz Khas, New Delhi 110016, India

[6]Cambridge Graphene Centre, Department of Engineering, University of Cambridge, 9 JJ Thomson Ave, Cambridge, CB3 0FA UK

[7]Center for Functional Nanomaterials, Brookhaven National Laboratory, Upton, NY 11973

[8]Department of Chemical Engineering and Biotechnology, Philippa Fawcett Drive, Cambridge, CB3 0AS United Kingdom

# These authors have contributed equally to the work.
e-mail: hs220@cam.ac.uk, sds65@cam.ac.uk

# Abstract

Investigation of the inherent field-driven charge transport behaviour of 3D lead halide perovskites has largely remained a challenging task, owing primarily to undesirable ionic migration effects near room temperature. In addition, the presence of methylammonium in many high performing 3D perovskite compositions introduces additional instabilities, which limit reliable room temperature optoelectronic device operation. Here, we address both these challenges and demonstrate that field-effect transistors (FETs) based on methylammonium-free, mixed-metal (Pb/Sn) perovskite compositions, that are widely studied for solar cell and light-emitting diode applications, do not suffer from ion migration effects as their pure Pb counterparts and reliably exhibit hysteresis free p-type transport with high mobility reaching 5.4 cm$^2$/Vs, ON/OFF ratio approaching 10$^6$, and normalized channel conductance of 3 S/m. The reduced ion migration is also manifested in an activated temperature dependence of the field-effect mobility with low activation energy, which reflects a significant density of shallow electronic defects. We visualize the suppressed in-plane ionic migration in Sn-containing perovskites compared to their pure-Pb counterparts using photoluminescence microscopy under bias and demonstrate promising voltage and current-stress device operational stabilities. Our work establishes FETs as an excellent platform for providing fundamental insights into the doping, defect and charge transport physics of mixed-metal halide perovskite semiconductors to advance their applications in optoelectronic devices.

## Introduction

Metal halide perovskites have earned the distinction of one of the most exciting emerging semiconductor technologies in the last decade, with photovoltaic power conversion efficiencies demonstrating a significant surge from 3.8% in 2009[1] to 25.5% today.[2] This achievement largely stems from the excellent optoelectronic properties of halide perovskites, including high absorption coefficient,[3] facile tunability of bandgaps,[4] long carrier diffusion length[5] and high tolerance to defects[6] among many others. 3D halide perovskites are generally described by an $ABX_3$ stoichiometry, where A is a monovalent organic/inorganic cation (e.g., methylammonium, $MA^+$; formamidinium, $FA^+$; cesium, $Cs^+$ etc.), B is a divalent metal cation (e.g. $Pb^{2+}$, $Sn^{2+}$ etc.) and X is a halide anion ($Cl^-$, $Br^-$, $I^-$ etc.). The domino effects of unprecedented developments in photovoltaics have been felt in other related emerging applications, including light emitting diodes,[7-9] lasers,[10,11] photodetectors,[12] and radiation detectors.[12,13] In particular, mixed metal (Pb/Sn) perovskite compositions, which are a unique class of perovskite materials as they allow realising low bandgaps that can't be reached with pure Pb compositions[14], have recently found increasing relevance in a range of optoelectronic devices, including all-perovskite multijunction solar cells[15,16] and near-infrared perovskite light emitting diodes[17]. While their optoelectronic properties have been studied in great depth, a detailed understanding of the charge transport physics of these mixed metal systems is still lacking.

FETs are three-terminal devices that probe the movement of charges at the interface of a semiconducting channel with the dielectric, which can be modulated with voltages applied at the gate and the drain terminals (both with respect to the source terminal). Unlike spectroscopic measurements which probe more localized transport and thereby provide an upper bound on the mobility obtained for a semiconductor,[18] FETs yield more realistic estimates of long-range transport that include the influence of interfaces and morphology, among others. In addition to determining the field effect mobility, FETs allow precise control of charge density to probe the trap energetics and instabilities of charge transport physics of both organic[19] and inorganic semiconductors[20,21].

Since the first report on halide perovskite-based FETs in 1999,[22] where Kagan et al. used a 2D perovskite $PEA_2SnI_4$ (PEA = phenethylammonium, $C_6H_5C_2H_4NH_3$) film as the semiconducting channel, subsequent attempts have largely focused on improving the performance of such 2D perovskite FETs through better processing protocols and architecture engineering of FETs.[22-28] In terms of 3D perovskites, compositional engineering at the A-site as well as halide modification have been shown to significantly affect the nature of charge transport in metal halide perovskite based FETs[29-32]. In this context, we have previously demonstrated that modification of the A-site cation resulted in enhancement of n-type device performance and improved operational stability in the Pb-iodide based perovskites[33]. Similarly, substitution of the halide anion from iodide to bromide resulted in modification of the nature of charge carriers from n-type to p-type respectively[30]. However, due to the soft nature of halide perovskites, undesirable ionic migration effects have largely hindered room-temperature charge transport studies on the more versatile 3D perovskite compositions which require careful optimisation of materials composition and processing.[29,31-35] Mobile ions in perovskite screen the applied potential and reduce the gate modulation of carriers at the interface, thereby leading to an apparent lower mobility at room temperature along with pronounced hysteresis and non-idealities in the device characteristics.[29,34,36] Moreover, most of the 3D perovskite FETs contain methylammonium cations (MA$^+$, CH$_3$NH$_3^+$), a species which is known to be inherently thermally unstable[37] and can introduce additional problems of dipolar disorder which further reduces the carrier mobility near room temperature[38]. Both these effects currently hamper the investigation of inherent charge transport behaviour in 3D halide perovskites using FET measurements.

In this work, we address the above challenges by exploring Pb-Sn alloying in mixed formamidinium-cesium (FA-Cs) perovskite compositions and employing them in bottom-gate, bottom-contact (BGBC) FET architectures (see **Methods**). These devices exhibit near ideal FET performance with tunable p-type mobility which allows us to probe the inherent charge transport mechanism in this class of perovskites. For the optimal mixed composition Cs$_{0.15}$FA$_{0.85}$Pb$_{0.5}$Sn$_{0.5}$I$_3$, we achieve a high hole mobility of 5.4 cm$^2$/Vs at room temperature, which is among the highest reported p-type field effect mobilities for 3D perovskite thin film FETs (**Table S1**). Temperature-dependent transport measurements reveal an interesting reversal from a negative temperature coefficient of mobility for pure-Pb compositions to a positive temperature-coefficient of mobility for optimized mixed Pb-Sn compositions. Such temperature-activated mobility behaviour is characteristic of inherent shallow defects prevalent in these Sn-containing halide perovskite semiconductors or factors such as B-site disorder which may modify the energetic landscape for charge transport[39]. We directly visualize the mitigation of ionic migration effects in Sn-containing perovskite compositions using microscopic photoluminescence measurements, and correlate these findings to the high mobility and operational stability of these perovskite devices. Our focus in this work is decidedly not to develop new materials systems that provide particularly high mobilities for FET applications. Instead the aim of our work is to advance the fundamental understanding of the key factors that govern the charge transport physics of low bandgap mixed metal (Pb/Sn) perovskite compositions that are actively being developed for optoelectronic applications including all-perovskite tandem solar cells and near-infrared LEDs.[38]

## Results & Discussion

Pure Sn-based 3D perovskites were initially chosen as candidate semiconductors for p-type perovskite FETs[39-41]. BGBC FETs were fabricated from thin films of pure Sn- based perovskites (such as $MASnI_3$, $FASnI_3$, $Cs_{0.15}FA_{0.85}SnI_3$) by solution processing on lithographically patterned Si/$SiO_2$ substrates which exhibited a clear signature of heavy p-type doping with no field effect modulation (**Figure 1a, S1**). This is consistent with previous reports of a high level of p-type self-doping arising from shallow Sn vacancies with low defect formation energy[14]. We observe that the bulk conductivity decreases when the A-site cation is modified from $MA^+$ to $(Cs_{0.15}FA_{0.85})^+$ (**Figure 1a**). We then systematically explored the mixed series of Pb/Sn compositions with $(Cs_{0.15}FA_{0.85})^+$ as the A-site cation to fabricate high-performance p-type perovskite FETs. Phase pure films were ascertained from detailed optical and structural analysis through the observation of the characteristic monotonic band bowing feature[42,43] (**Figure S2**) and crystalline peaks corresponding only to the perovskite phase[44] (**Figure S3**). Similar structural analysis performed with A-site variation while retaining the B-site composition of $Pb_{0.5}Sn_{0.5}$ also exhibited clean phase pure crystalline peaks of perovskite (**Figure S4**); see later sections for further discussion on the transport behaviour of A-site variant perovskite compositions.

**Figure 1** shows the device characteristics of BGBC FETs fabricated from different perovskite compositions with varying Pb/Sn ratios. To begin with, FETs were fabricated with the more conventional Pb-based $Cs_{0.15}FA_{0.85}PbI_3$ (referred to hereafter as CsFAPbI3) composition. These devices exhibited an n-type field effect transport (**Figure 1a, S5**) with a low $\mu_{FET} \sim 10^{-3}$ cm$^2$/Vs, similar to our earlier studies[33]. Interestingly, upon substitution of Pb$^{2+}$ with Sn$^{2+}$ even by 25% (Cs$_{0.15}$FA$_{0.85}$Pb$_{0.75}$Sn$_{0.25}$I$_3$), we observe a clear transition from n-type to p-type field effect transport (**Figure 1b, S6**) and the room-temperature hole mobility increases to ~0.02 cm$^2$/Vs. However, both these perovskite compositions exhibit notable hysteresis as evident from the transfer (**Figure S5 & S6**) and output (**Figure S7**) characteristics, which bear signature of ionic defect migration in an operational FET[29,34]. Further increasing the Sn content to 50% (Cs$_{0.15}$FA$_{0.85}$Pb$_{0.5}$Sn$_{0.5}$I$_3$, referred to hereafter as CsFAPb$_{0.5}$Sn$_{0.5}$I$_3$) increases the channel current by three orders of magnitude and $\mu_{FET}$ reaches a maximum of 5.4 cm$^2$/Vs (average $\mu_{FET} \sim 3.1$ cm$^2$/Vs measured on 12 devices over 3 batches) (**Figure 1c & S8**). We note that the reliability factor for mobility estimation is relatively high (~80-100 % and in some cases has values greater than 100 %)[45] demonstrating that mobility estimations are robust and conservative (**Figure S9-S10**). The mobility variation of FETs fabricated upon B-site (Pb/Sn) variation is presented in **Figure 1c**. Importantly, devices fabricated from CsFAPb$_{0.5}$Sn$_{0.5}$I$_3$ also exhibit clean hysteresis-free output characteristics with a well-defined linear and saturation regime (**Figure 1d**). Such clean hysteresis-free behaviour has been rare and difficult to achieve in the many reported 2D and/or 3D perovskite FET studies in the literature, but with appropriate compositional tuning, we have been able to reduce the effects of ionic defects to an extent that clean textbook-like transistor characteristics are obtained. Moreover, these devices exhibited excellent FET characteristics including gate-induced current modulation with ON/OFF ratio approaching 10$^6$ and channel conductance of 3 S/m, which are among the best in the field of perovskite FETs (**Figure S11** and **Table S1**). It is noteworthy that in comparison to Pb-based perovskites which exhibit switch on voltage close to 0 V, these Sn based perovskites exhibit rather large positive turn on voltage of ~ 20 - 30 V which corresponds to a p-type background doping of ~ 10$^{12}$ cm$^{-2}$. Further increase of the Sn fraction to 75% with a perovskite composition of Cs$_{0.15}$FA$_{0.85}$Pb$_{0.25}$Sn$_{0.75}$I$_3$ (referred to hereafter as CsFAPb$_{0.25}$Sn$_{0.75}$I$_3$) results in a drop in gate modulation (I$_{ON}$/I$_{OFF} \sim 1.5$) due to the overwhelming extent of p-type self-doping (corresponding p-type background doping ~10$^{14}$ cm$^{-2}$) giving rise to quasi-metallic type conduction (**Figure S12**). **Table S2** summarizes the key FET parameters obtained for various Pb-Sn mixing ratios.

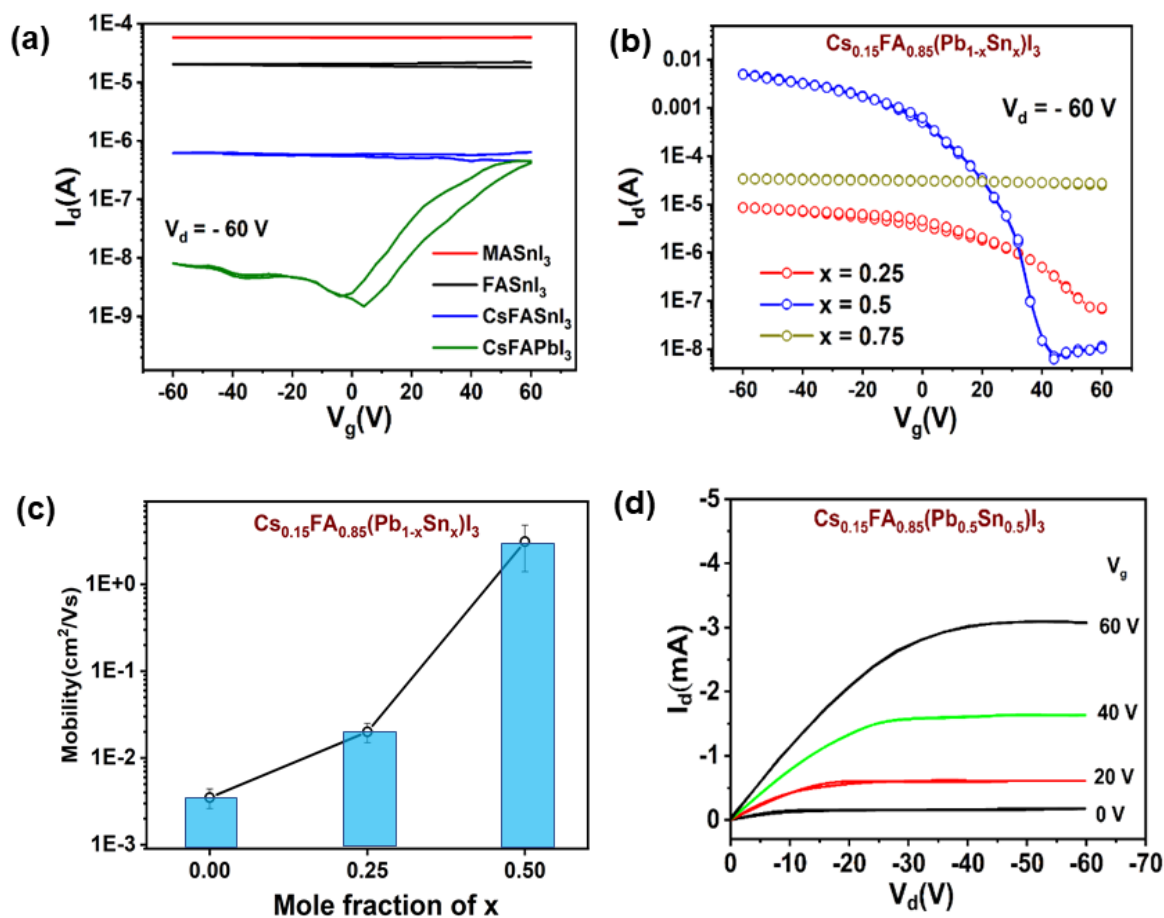

**Figure 1: FET characterization on Pb-Sn perovskite films.** Transfer characteristics measured on typical BGBC FETs (L = 40 µm, W = 1mm) fabricated from different perovskite compositions, (a) $MASnI_3$, $FASnI_3$, $Cs_{0.15}FA_{0.85}SnI_3$ and $Cs_{0.15}FA_{0.85}PbI_3$ (b) $Cs_{0.15}FA_{0.85}Pb_{1-x}Sn_xI_3$ with x = 0.25, 0.5 and 0.75. (c) Saturation field effect mobility distribution upon variation in x from 0 to 0.5. (d) Output characteristics measured on $Cs_{0.15}FA_{0.85}Pb_{0.5}Sn_{0.5}I_3$ perovskite FETs at 300K.

To understand the transport trends upon B-site compositional variation, we performed *ab initio* electronic structure calculations (see details of calculations in **Methods**). To construct a computationally tractable supercell without deviating much from the experimental one, we adopted the A-site composition to be $Cs_{0.125}FA_{0.875}$ and the relative composition of Pb/Sn was varied systematically (See **Methods IV** for details). In Pb-iodide perovskites, the 6s-orbitals of Pb and 5p-orbitals of iodine contribute to the valence band (VB) edge, whereas the conduction band (CB) edge has dominant contributions from Pb 6p-orbitals with only weaker influence from 5p-orbitals of iodine (**Figure 2a-b, S13a-b**)[46,47]. However, upon introduction of $Sn^{2+}$, the band decomposed charge densities and partial density of states (pDOS) plots depict that the Sn 5p-orbitals dominantly contribute to the VB edge of all the B-metal alloyed compositions (**Figures 2a-d & S13**). The electronic contributions from Pb 6s-orbitals and I 5p-orbitals to the VB edge remain secondary for these mixed lead-tin iodide perovskites. Nevertheless, the CB edge is dominated by the 6p–orbitals of Pb atoms for all systems and has relatively minimal contribution from the Sn atoms. Overall, B-metal mixing results in enhanced density of states predominantly at the VB edge. We further investigate the relative energetic positions of band edges to qualitatively understand the hole transport in these materials. For this, we employ a computational band alignment scheme that has been widely used for similar perovskites[47-49]. This analysis reveals that the valence band maxima (VBM) shift upward in energy as the Sn concentration increases (**Figure S13**) which is expected to lead to improved hole injection.

In semiconductors, the carrier effective masses at the VBM and conduction band minima (CBM) are crucial for understanding the change in the charge carrier mobilities with compositional engineering. We computationally calculate the hole and electron effective masses from the band curvature of edge-states at the VBM and the CBM, respectively. As the effective mass and corresponding band curvature are inversely related, the higher band-edge dispersion represents lighter effective mass. The calculated effective masses in $CsFAPbI_3$ reveal that holes are heavier ($m_h^* = 0.19$ $m_e$) in the VBM compared to electrons ($m_e^* = 0.13$ $m_e$) in the CBM (**Figure S13d**). Upon Pb/Sn alloying, the reduced effective mass ($m_e^* m_h^* / (m_e^* + m_h^*)$) exhibits a monotonic decrease (0.077-0.052 $m_e$, **Figure 2e**) with an increase in Sn contents, in line with the earlier observation from magneto-optical measurements[50]. To pin down the origin of the variation in the effective mass with B-metal mixing, we investigate the change in B-I-B bond angles of the lattice. **Figure 2f** exhibits a systematic enhancement in the average B-I-B bond angle with the increase of Sn fraction in the computational system $Cs_{0.125}FA_{0.875}Pb_{(1-x)}Sn_xI_3$. This geometric change can be ascribed to the fact that upon substitution of Pb atoms with Sn, the crystal structure changes from a slightly distorted tetragonal phase to a less distorted cubic-like phase[43,51,52]. With the increased linearity in B-I-B angles, the s-orbitals of B-atom and 5p-orbitals of I enhance their spatial overlap (**Figure 2g**). Such an increase in the overlap between the participating orbitals boosts the dispersion at the VBM state and thereby reduces the hole effective mass. Furthermore, the morphological grain size obtained from our optimized processing condition consistently increases from ~ 106 nm for CsFAPbI$_3$ to ~ 150 nm, ~ 275 nm and ~ 401 nm for CsFAPb$_{0.75}$Sn$_{0.25}$I$_3$, CsFAPb$_{0.5}$Sn$_{0.5}$I$_3$ and CsFAPb$_{0.25}$Sn$_{0.75}$I$_3$, respectively (**Table S3 and Figure S14**). Thus, upon addition of Sn, the reduction of hole effective mass, together with the increase of mean morphological grain size, results in the observed high hole mobility in the mixed Pb-Sn FET devices.

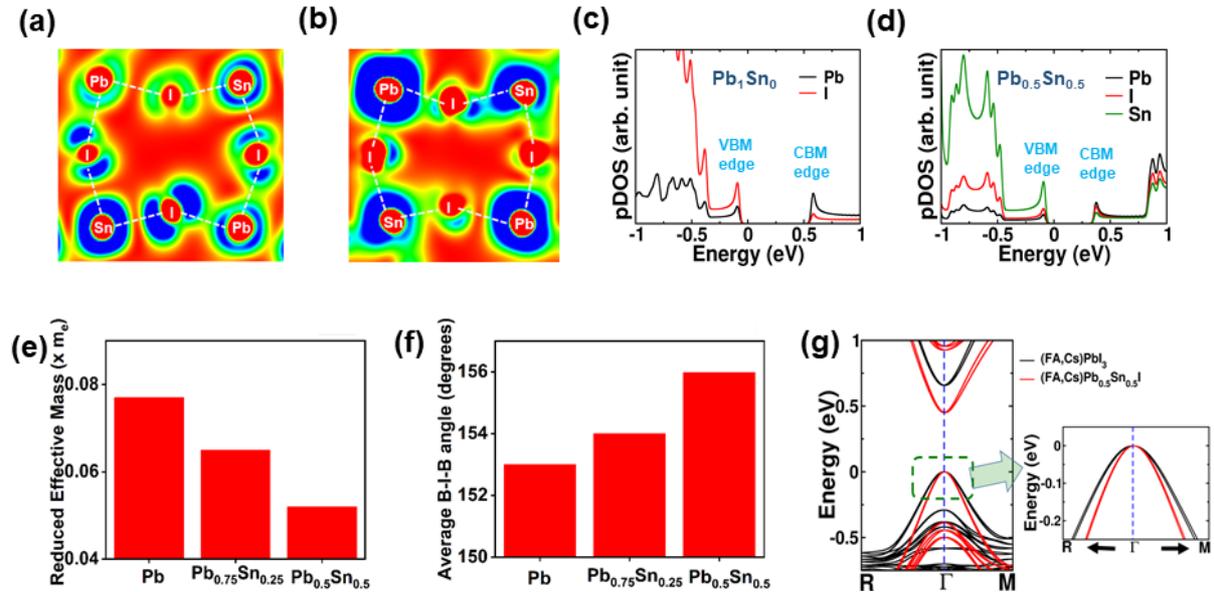

**Figure 2: Atomistic origin of high mobility p-type transport in mixed Pb-Sn devices.** Electronic charge density contours corresponding to the (a) VBM and (b) CBM of CsFA Pb$_{0.5}$Sn$_{0.5}$I$_3$. Color scale: red is defined as 0 and blue as 0.00012 eÅ$^{-3}$. Plots of partial density of states for (c) CsFAPbI3 and (d) CsFA Pb$_{0.5}$Sn$_{0.5}$I$_3$. (e) Variation of hole effective mass with different Pb/Sn mixing ratios (y) in CsFAPb$_{(1-x)}$Sn$_x$I$_3$. (f) Average B-I-B bond angles with the variation of Sn contents in the lattice. (g) Band structure of CsFAPbI$_3$ and CsFAPb$_{0.5}$Sn$_{0.5}$I$_3$, normalized at the VBM. Zoomed versions of their valence band edges have also been shown to indicate differences in band curvatures near the VBM.

Sn has two oxidation states of +2 and +4, with the latter being more thermodynamically stable, whereas an oxidation state of +2 is essential for the ABX$_3$ perovskite structure. Such facile oxidation of Sn$^{2+}$ to Sn$^{4+}$ in the presence of traces of oxygen or typical solvents (e.g. dimethyl sulfoxide, DMSO) at high temperature causes unwanted performance degradation in devices and affects their long-term stability[14]. In order to ascertain the relative abundance of Sn$^{2+}$ and Sn$^{4+}$ in the deposited films, we performed X-ray photoelectron spectroscopy (XPS) measurements on perovskite thin films with varied Sn contents in the composition, fabricated on the same substrates (without Au electrodes) utilized for fabricating the FETs (see survey spectra in **Figure S15** and Pb 4f and I 3d high-resolution spectra in **Figure S16)**. Since Sn atoms play a significant role in dictating the p-type charge transport in our films, we focus our analysis on the XPS spectra corresponding to Sn $3d_{5/2}$ core-levels. All the perovskite thin films exhibited two characteristic peaks indicating the co-existence of $Sn^{4+}$ and $Sn^{2+}$ (**Figure 3a-d**). The spectral feature at binding energy of 486.40 eV corresponds to $Sn^{2+}$, whereas the spectral feature at a slightly higher binding energy of 487.57 eV is distinguished as $Sn^{4+}$.[53] We estimate the relative abundance of the two oxidation states $\frac{Sn^{4+}}{Sn^{2+}}$ (estimated from the area under the peaks at the binding energy of the respective oxidation states), which is plotted in **Figure 3e**. It is evident that our champion composition of 50% Sn has the lowest ratio of 0.06. Interestingly, the 25% Sn composition shows the largest ratio of 3.74, which is even higher than that of 100% Sn (1.35). This result is in agreement with a recent report showing that at low Sn-contents of 20-40%, the resulting thin films are prone to increased defect densities and exhibit inferior photovoltaic performance[54]. We see a qualitatively similar trend with increasing $\frac{Sn^{4+}}{Sn^{2+}}$ corresponding to increased Urbach energy ($E_u$) (**Figure 3e**), a quantity representing band edge disorder which is obtained from photothermal deflection spectroscopy (PDS) measurements (**Figure S17**), thereby suggesting a direct correlation between $Sn^{4+}$ and defect activity.

Since defects are a major source of non-radiative recombination in semiconductors, we sought to examine the effect of Sn content in the perovskite composition on the integrated photoluminescence (PL) output from the fabricated thin films of similar thickness (~ 200 nm). Interestingly, we observe that the PL count increases with increase in Sn content in the overall perovskite composition, irrespective of the fraction of $Sn^{4+}$ (**Figure 3f**). This suggests that the increase in background doping upon increasing the relative Sn composition in the perovskite possibly dominates the PL emission in these Sn-containing perovskites, especially for Sn contents >50%. At the same time, shallow Sn vacancies which are considered to be the real cause of such background doping may influence the $E_u$.

We explored the dependence of charge transport behaviour on composition by using various combinations of $FA^+$, $Cs^+$ and $MA^+$ at the A-site, while maintaining the B-site composition at $Pb_{0.5}Sn_{0.5}$ and iodide as the halide ion (see summary in **Table**

**S4**). FET devices fabricated from FAPb$_{0.5}$Sn$_{0.5}$I$_3$ exhibit a p-type field effect transport with a μ$_{FET}$ ~ 0.15 cm$^2$/Vs (**Figure S18**). Upon incorporation of Cs$^+$ cation in various amounts (5%, 15%, 25% and 40%) to replace FA$^+$, we observe that the optimum Cs content to obtain the highest mobility and hysteresis free transfer characteristics is 15% (**Figure S18b-f, S19-S21**), as we employed earlier in this work.

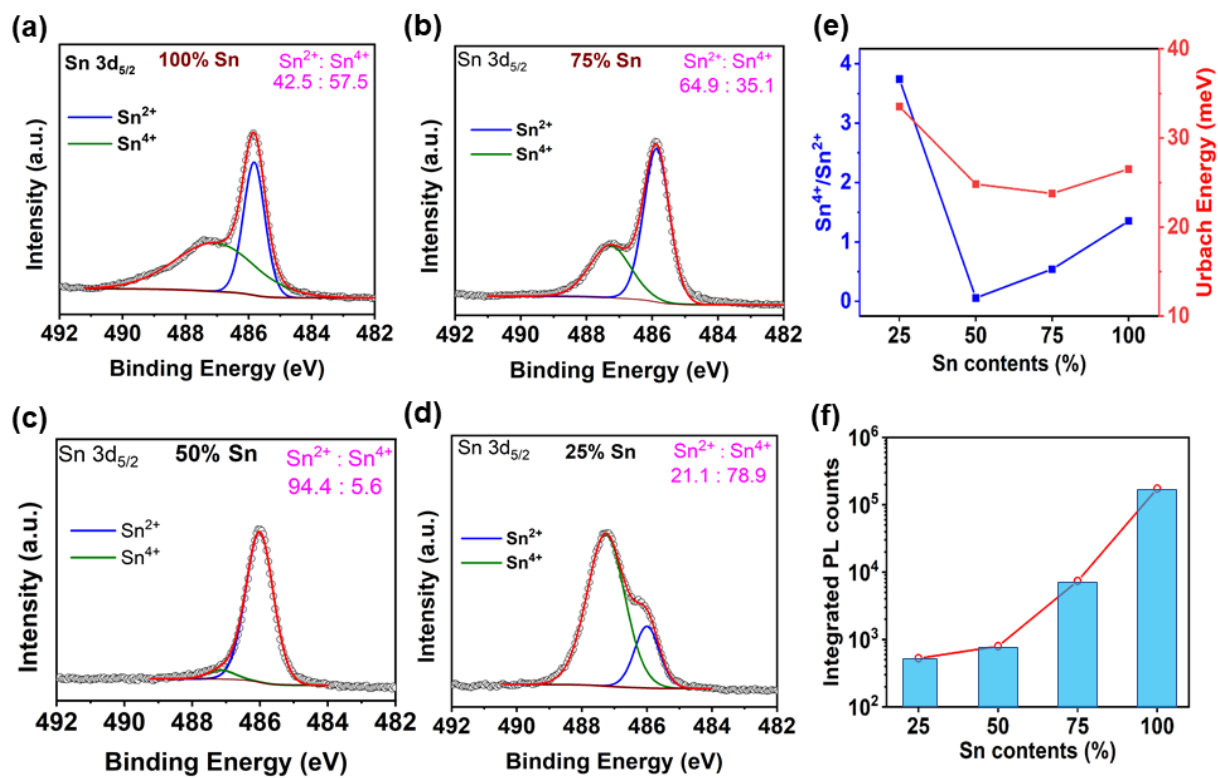

**Figure 3: Chemical analysis of defects in mixed Pb-Sn perovskite films.** (a-d) High resolution XPS spectra of Sn 3d$_{5/2}$ core levels performed on perovskite compositions with different Pb/Sn mixing ratios. Brown line is the background, red line is the best-fitting to the raw data. The relative abundance (%) of Sn$^{2+}$ and Sn$^{4+}$ in the thin films are determined using suitable fitting. (e) Plot of the ratio of Sn$^{4+}$/Sn$^{2+}$ for different Pb-Sn mixing ratios (blue), shown together with the Urbach energy obtained from PDS measurements (red). (f) Variation of integrated PL counts for perovskite films with different Pb/Sn compositions while maintaining the film thickness.

Similarly, using MA$^+$ (instead of Cs$^+$) to partially substitute FA$^+$ in FAPb$_{0.5}$Sn$_{0.5}$I$_3$ also results in an increase in mobility up to 0.8 cm$^2$/Vs for 40% MA contents (**Figure S22-S24**). However, at the same time, the hysteretic behaviour increases monotonically with the increase in MA$^+$ composition (**Figure S22-S24**), which bears a signature of dipolar disorder originating from the polar MA$^+$ cation.[29,33] The increase in mobility for dual cation mixtures (CsFA or MAFA) relative to the pristine FA composition is in agreement with the enhancement in crystallinity observed from the increase in peak counts and sharpening of the XRD patterns (**Figure S4**). Morphological grain size extracted from top-view SEM measurements increased from 192 nm for $FAPb_{0.5}Sn_{0.5}I_3$ to 241 nm and 250 nm for optimized compositions of Cs (15%) and MA (40%) respectively (**Figure S25** and **Table S5)**. Further optimization procedures such as treatment of contacts with self-assembled monolayers[29] (**Figure S26, S27**) or modification of the perovskite precursor solution with additives such as diethyl sulphide (DES)[55] did not exhibit any improvement in FET performance (**Figure S28**).

To understand the charge transport mechanisms in these mixed-metal perovskite compositions, we investigated the temperature-dependent transport for FETs fabricated with different Pb/Sn mixing ratios. Moreover, we compared $\mu_{FET}(T)$ obtained from forward and reverse sweeps specifically for the devices with hysteretic characteristics. Mobility extracted from forward and reverse sweeps in some of the hysteretic perovskite compositions vary marginally from 5 – 7 % over the complete temperature range. For the plots in **Figure 4,** we used the values of $\mu_{FET}$ obtained from the forward characteristics, which tends to be the more conservative estimate. **Figure S29** depicts detailed temperature-dependent transfer characteristics for different perovskite compositions. FETs fabricated from $CsFAPbI_3$ exhibit a clear negative coefficient of mobility over the complete temperature range (**Figure 4a**), which is consistent with our earlier observation on pure-Pb based perovskite FETs.[29,33] This negative coefficient of mobility reflects the temperature dependence of ionic migration and the associated screening of the gate field by negatively charged halide ions. In contrast, FETs fabricated from $CsFAPb_{0.75}Sn_{0.25}I_3$ exhibit a positive coefficient of mobility (**Figure 4b**) indicating that thermally activated processes limit the charge transport in these systems and that the lowering of mobility due to ion migration is substantially suppressed with the incorporation of Sn. In fact, such a temperature-activated trend has, to the best of our knowledge, never been reported in 3D hybrid perovskite FETs. In the temperature range between 200 K and 300 K, an Arrhenius-like dependence is observed, from which the activation energy ($E_A$) is extracted to be 160 meV and for T < 200 K, $E_A$ ~ 57 meV is obtained (**Figure 4b**). Upon increasing the Sn content to 50% (**Figure 4c**), $E_A$ lowers to a value of 48 meV in the range of 200 K to 300 K and decreases to as low as ~ 10 meV for T < 200 K. Such a low value of $E_A$ is indicative of inherent shallow traps prevalent in this class of Pb-Sn based solution-processed perovskite semiconductors and is consistent with the high values of μ<sub>FET</sub> observed in the FETs. In fact, Sn vacancies, which are rife in Sn-containing perovskites and believed to be responsible for the unintentional background doping are known to be shallow and hence these could be a source of such shallow traps characteristic with low $E_A$ values.[39] Another source of such activated behaviour could be the B-site disorder possibly present in these systems, which may result in fluctuation in the energetic landscape of the VB, thereby affecting the p-type transport.

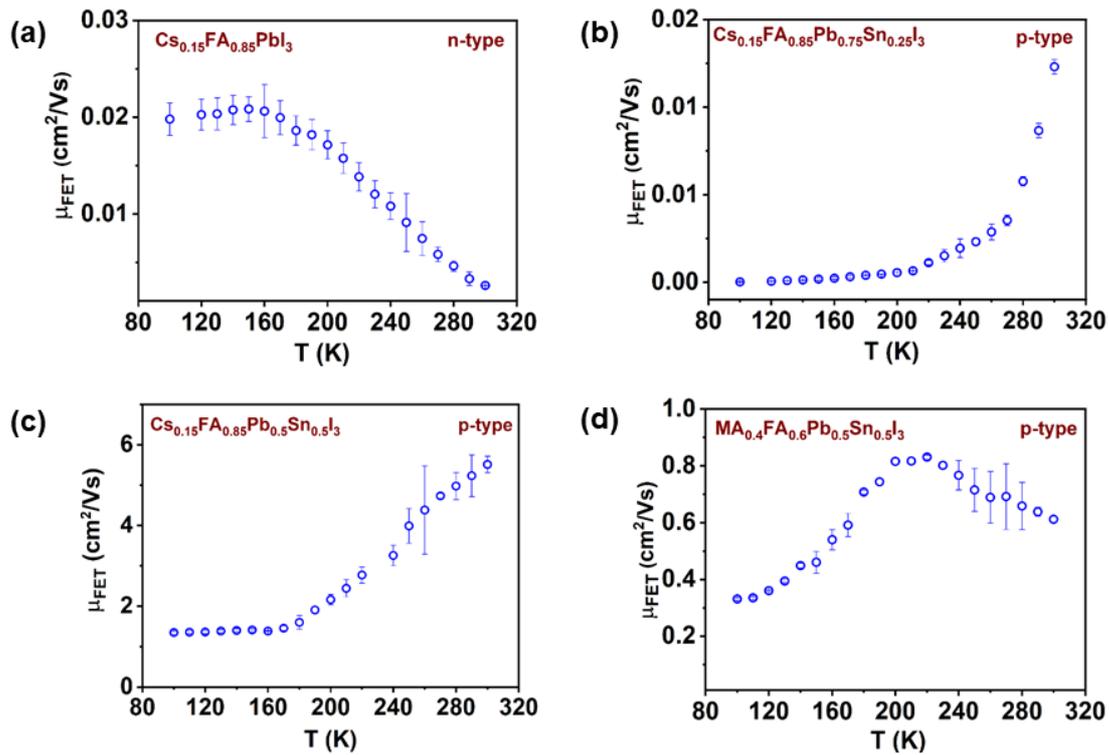

**Figure 4: Temperature dependent charge transport.** μ<sub>FET</sub> (T) measurement performed on typical BGBC FETs (L = 40 μm, W = 1mm) for different perovskite compositions. (a) $Cs_{0.15}FA_{0.85}PbI_3$, (b) $Cs_{0.15}FA_{0.85}Pb_{0.75}Sn_{0.25}I_3$, (c) $Cs_{0.15}FA_{0.85}Pb_{0.5}Sn_{0.5}I_3$ and (d) $MA_{0.4}FA_{0.6}Pb_{0.5}Sn_{0.5}I_3$.

μ<sub>FET</sub>(T) measurements on FETs fabricated from $MA_{0.4}FA_{0.6}Pb_{0.5}Sn_{0.5}I_3$ exhibit a different behaviour (**Figure 4d**). In these devices, the mobility follows an activated temperature dependence over the temperature range of 100-200 K. However, as the temperature increases >200 K, μ<sub>FET</sub>(T) exhibits a negative coefficient. Moreover, the magnitude of the transition temperature increases to ~ 220 K (**Figure S30**) for FETs fabricated with lower MA contents ($MA_{0.15}FA_{0.85}Pb_{0.5}Sn_{0.5}I_3$), thereby indicating that this is a characteristic feature of MA containing Pb-Sn perovskite-based FETs. In the case of MA[+], the polar nature of the A-site cation forms H-bonding with the octahedral cage thereby destabilizing the lattice structure and generating significant ionic defects which we had earlier shown in our report[29] to reduce the field effect mobility in the similar temperature range.

In order to ascertain whether the trend of $\mu_{FET}(T)$ observed in our mixed Pb-Sn perovskite FETs is related to any crystal phase transitions, we performed temperature dependent XRD measurements on our champion $CsFAPb_{0.5}Sn_{0.5}I_3$ thin films (**Supplementary Note 13**). We notice a possible phase transition region between 122 K and 182 K for (111) peak (**Figure S31-S32**), where multiple peaks start appearing at 172 K and they show complex changes until 122 K when a single and very broad peak reappears indicating transition to a lower symmetry phase. It is important to note that the structural variation as observed in the said temperature range is not abrupt and is likely also complicated by the polycrystalline nature of the films. Interestingly, the onset of such structural changes is close to the temperature where $\mu_{FET}(T)$ also begins to rise with temperature (**Figure 4c**). However, the nature of this structural transition is unclear in this current work and further work would be required to better understand this relationship with transport.

In order to visualize the in-plane ionic migration under bias in these mixed perovskite thin films, we show photoluminescence (PL) mapping results (in nitrogen atmosphere) on lateral devices (L ~ 20 μm) in **Figure 5**. Irrespective of the perovskite composition, unbiased devices exhibit uniform PL intensity throughout the channel. Devices were then subjected to multiple cycles of poling (30 V for 30 seconds). Upon biasing, the PL intensity profile in the channel area continuously evolves for $CsFAPbI_3$, which can be correlated with halide migration upon biasing, as shown in our previous work (**Figure 5a, S33-S34**)[33]. Due to accumulation of excess halide ions near the positive electrode, possibly in the form of iodide interstitials, reduction of PL happens because they are known to act as deep traps in Pb perovskites[56] and hence increases non-radiative recombination. Such PL reduction then gradually percolates through the entire channel upon subsequent biasing cycles. Similar measurements were also performed on lateral devices fabricated from $CsFASnI_3$ (**Figure 5b, S35**), $CsFAPb_{0.25}Sn_{0.75}I_3$ (**Figure S36**) and $CsFAPb_{0.5}Sn_{0.5}I_3$ (**Figure 5c, S37**). However, despite multiple biasing cycles, the PL profiles in the channel area of Sn-containing perovskites do not exhibit significant variation (**Figures 5a-c, S35-37**) except for the initial variation of PL profile in the channel after the first bias, which could point to the bias dependent re-organization of $Sn^{4+}$ in the channel. Interestingly, the PL intensity over the negative electrode monotonically increases with bias. Furthermore, this process of photo-brightening of the negative electrode upon biasing appears to be a reversible process and the PL intensity over the negative electrode reverts to its original intensity when left unbiased for an extended period in the dark (**Figure 5d, S38-39**). Such enhancement of PL intensity on the negative electrodes is not observed in $CsFAPbI_3$ (**Figure S33**) and is specific to the Sn-containing compositions. The relative invariance of the PL profiles in the channel area upon biasing indicates that bias induced lateral ionic migration process is significantly minimized in Sn-containing perovskites (**Figure 5e**), which is responsible for the outstanding transport properties described earlier.

To understand the bias induced PL enhancement on the negative electrode, we plot the normalized PL counts for different perovskite compositions as a function of total biasing duration (**Figure 5f**). These plots exhibit a clear characteristic of electrochemical processes and considering that this process occurs primarily on the negative electrode, we propose that this enhancement is indicative of the local electrochemical reduction of $Sn^{4+}$ to $Sn^{2+}$. Since $Sn^{4+}$ has earlier been described as a major source of defects in Sn-containing perovskites, the increase in radiative emission at the negative electrode upon biasing can be attributed to the reduction in $Sn^{4+}$ defect densities. We estimated the characteristic time scales of this electrochemical process for different Sn-containing perovskite compositions and found that the rate constant for the reduction of $Sn^{4+} \rightarrow Sn^{2+}$ is at least 200 times slower for $CsFAPb_{0.5}Sn_{0.5}I_3$ in comparison to $CsFASnI_3$ (**Figure 5f**). From Le Chatelier's principle, this decrease in rate constant can be correlated to the decrease of effective $Sn^{4+}$ concentration in $CsFAPb_{0.5}Sn_{0.5}I_3$ as compared to that in $CsFASnI_3$, which agrees well with the XPS measurements described earlier. As the brightening of the negative electrode is reversible after turning off the bias (**Figures 5d and 5g**), we attribute the subsequent lowering in PL during relaxation to the spontaneous oxidation of $Sn^{2+}$ to $Sn^{4+}$. All in all, such precise control of defect dynamics in Sn-containing perovskites through compositional changes upon biasing brings out the transformative nature of perovskites towards designing high performance electronic devices.

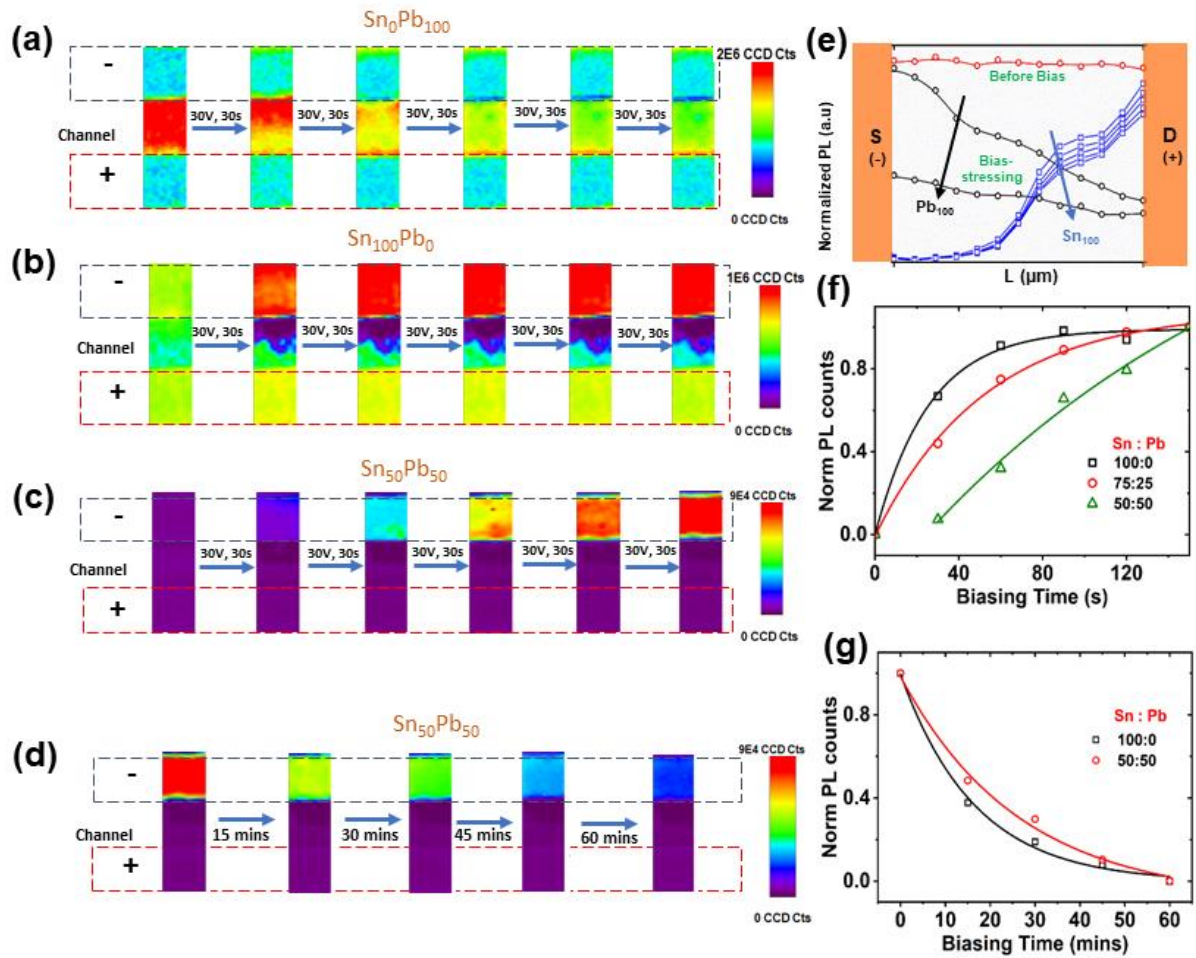

**Figure 5: Lateral ion migration in perovskites.** Photoluminescence (PL) mapping performed on lateral channel devices (L = 20 µm) of (a) $Cs_{0.15}FA_{0.85}PbI_3$, (b) $Cs_{0.15}FA_{0.85}SnI_3$ and (c) $Cs_{0.15}FA_{0.85}Pb_{0.5}Sn_{0.5}I_3$ perovskites upon multiple cycles of biasing (30V, 30 s). (d) Change in PL with time for $Cs_{0.15}FA_{0.85}Pb_{0.5}Sn_{0.5}I_3$ device kept in dark upon removal of bias. (e) Evolution of PL line profile of the channel area in lateral devices of $Cs_{0.15}FA_{0.85}PbI_3$ and $Cs_{0.15}FA_{0.85}SnI_3$ upon multiple bias stressing. Corresponding normalized PL counts obtained from the negative electrodes of devices with varying Pb/Sn mixing ratios upon (f) biasing and (g) removal of bias.

Having established that ionic migration effects are greatly suppressed in mixed Pb-Sn perovskite FETs, we tested the operational stability (observed from threshold voltage shifts $\Delta V_{th}$ with time) of fabricated devices with different A-site cations, viz. $FA^+$, $(Cs_{0.15}FA_{0.85})^+$ and $(MA_{0.4}FA_{0.6})^+$, while keeping the optimum B-site composition of $Pb_{0.5}Sn_{0.5}$ (**Figure S40**). It is clear that CsFA based composition imparts the best stability ($\Delta V_{th}$ ~ 4V for voltage-bias and ~ 5V for current-bias stress), while the presence of MA makes it unstable ($\Delta V_{th}$ ~ 25V for voltage-bias and ~ 32V for current-bias stress) over longer operation periods, thereby reinforcing the need to employ MA-free compositions in future. Besides operational stability, our champion composition of $Cs_{0.15}FA_{0.85}Pb_{0.5}Sn_{0.5}I_3$ also shows improved shelf-life stability over 2 months with a negligible drop in performance (devices stored in $N_2$ glovebox between successive measurements), which is far superior to both $FAPb_{0.5}Sn_{0.5}I_3$ and $FA_{0.6}MA_{0.4}Pb_{0.5}Sn_{0.5}I_3$ (**Figure S40**).

Finally, we propose the mechanism to account for the suppressed ionic migration effects observed in mixed metal perovskite compositions. In the case of Pb perovskite FETs, ionic migration (primarily by halides) screens the field-induced carriers from the positive gate potential and hence lowers their apparent mobility near room temperature. On the other hand, in mixed Pb-Sn perovskite FETs application of negative gate potential results in the accumulation of positively charged ions ($A^+$, $B^{2+}$) of the perovskite at the interface, which tend to be of lower ionic mobility than the halide defects in pure Pb-based perovskites. In addition, these defects are compensated by the presence of a large density of negatively charged Sn-vacancies ($V_{Sn}^{2-}$ that give rise to doping) resulting in two types of possible neutral ionic defect complexes i.e. ($2A^+$-$V_{Sn}^{2-}$) or ($B^{2+}$-$V_{Sn}^{2-}$) at the charge transport interface. This minimizes the ionic screening of the gate potential thereby allowing us to probe the inherent activated transport behaviour signifying residual shallow electronic traps. Furthermore, the optimized CsFA composition is free from $MA^+$ cation-induced dipolar disorder, which also contributes to the enhanced charge transport near room temperature. This represents a major leap in the field of 3D perovskite FETs which have until now been severely affected with undesirable ionic migration effects. Going forward, various molecular and interfacial doping strategies can be devised even for 3D Pb perovskites to inundate in a similar way such undesirable ionic screening effects and demonstrate inherent transport behaviour near room temperature.

## Conclusion

We systematically studied the charge transport physics of low bandgap 3D mixed-metal (Pb-Sn) perovskite semiconductors, which are optimal candidates for applications in all-perovskite tandem photovoltaics, near-infrared perovskite light emitters and photodetectors. Incorporation of Sn to partially replace Pb at the B-site modifies the electronic structure to obtain a reduced hole effective mass and enhanced density of states at the VB edge, thereby resulting in a pronounced p-type transport. Perovskite FETs fabricated from such mixed Pb-Sn perovskites exhibit excellent transport properties with μFET reaching as high as 5.4 cm$^2$/Vs, ON/OFF ratio approaching 10$^6$ and normalized channel conductance of 3 S/m. The mobility values are surprisingly high given the inevitable energetic disorder that is present in these mixed metal compositions. We also observe mixed-metal systems to be much less affected by ion migration than their pure Pb-based counterparts: Temperature-dependent transport measurements indicate a transition from an ionic migration-dominated negative coefficient of mobility to a temperature-activated regime, which is a manifestation of the inherent electronic traps that are present in solution-processed mixed-metal perovskite semiconductors. Such transport behaviour has, to the best of our knowledge, not been observed to date in 3D hybrid perovskite FETs, whose transport physics is normally governed by undesirable ionic screening effects, which we have successfully mitigated here. We were able to directly visualize the suppressed in-plane ionic migration and associated electrochemical processes occurring under bias and rationalize this behaviour within the framework of defect species that are present in these mixed-metal perovskites. Our work establishes FETs as a powerful and reliable platform for investigating the fundamental physics of doping, defects, instabilities, and charge transport in mixed metal halide perovskite semiconductors and in this way provides important insights furthering the progress of these materials in solar cells, LEDs and other optoelectronic applications.

## Acknowledgements

SPS acknowledges funding support from Royal Society through the Newton Alumni Fellowship (AL\211004, AL\201019, and AL\191021), SERB-SRG (SRG/2020/001641), DAE, Government of India. K.D. acknowledges the support of the Cambridge Trust and SERB (Government of India) in the form of Cambridge India Ramanujan Scholarship. K.D. thanks Dr. Felix Lang for useful discussions on photoluminescence measurements. This work has received funding from the European Research Council under the European Union's Horizon 2020 research and innovation programme (HYPERION, grant agreement no. 756962). R.H.F and R.S. acknowledge funding and support from the SUNRISE project (EP/P032591/1). R.S. acknowledges a Newton International Fellowship from The Royal Society. J.L.M-D. and W.-W. L. thank the UK Royal Academy of Engineering, grant CiET1819_24, EPSRC grants EP/N004272/1, EP/P007767/1, EP/L011700/1, The Winton

Programme for the Physics of Sustainability. W.-W. L. acknowledges Bill Welland for useful discussions. BR acknowledges the EPSRC, Grant Number EP/T02030X/1. NT would like to acknowledge using resources of the Center for Functional Nanomaterials (CFN), which is a U.S. Department of Energy Office of Science User Facility, at Brookhaven National Laboratory under Contract No. DE-SC0012704. DG is thankful to the Center for Integrated Nanotechnologies (CINT), a U.S. Department of Energy and Office of Basic Energy Sciences user facility, at Los Alamos National Laboratory (LANL), USA, for providing computational facilities. Y.Z. thanks the Chinese Scholarship Council and the EPSRC Centre for Doctoral Training in Graphene Technology for financial support. H.S. thanks the Royal Society for support through a Royal Society Research Professorship (RP\R1\201082). S.D.S. acknowledges support from the Royal Society and Tata Group (UF150033).

## Conflict of Interest

S.D.S. is a co-founder of Swift Solar, Inc.

## Methods

### I. Device fabrication

Typical bottom contact bottom gated devices were fabricated using lithographically patterned Au S-D electrodes on $Si/SiO_2$ substrates. This was followed by the deposition of the perovskite layers of varying compositions.

### II. Preparation of perovskite solutions

All the starting precursors e.g. Formamidinium Iodide (FAI, GreatCell Solar Materials), Methylammonium Iodide (MAI, GreatCell Solar Materials), Cesium Iodide (CsI, Sigma Aldrich, anhydrous, beads, −10 mesh, 99.999% trace metals basis), Lead Iodide ($PbI_2$, TCI), Tin Iodide ($SnI_2$, Sigma Aldrich, anhydrous, beads, −10 mesh, 99.99% trace metals basis) and Tin Fluoride ($SnF_2$, Sigma Aldrich) were dissolved in required ratios in a 3:1 (vol/vol) mixed solvent containing Dimethylformamide (DMF, Sigma Aldrich) and Dimethyl Sulfoxide (DMSO, Sigma Aldrich). The precursors were dissolved by shaking with hands and no heating or stirring was done during this process. The solutions were then left for 2-3 hours and then 5 vol% of formic acid (Sigma Aldrich, 95%) were added to each of them. Perovskite solutions were filtered through a 0.2 um

PTFE filter prior to spin coating. Preparation of all the solutions and their subsequent spinning were performed in a $N_2$-filled glovebox with $H_2O$ levels typically < 0.1 ppm and $O_2$ levels usually < 1 ppm.

## III.  Fabrication of perovskite films

Within one hour of adding formic acid, BGBC transistor substrates were treated with $O_2$ plasma for 2-3 minutes and then quickly transferred inside the glovebox for the deposition of perovskite films. 30 uL of perovskite solution was uniformly introduced on the substrate and spun at 5000 rpm (acceleration ~ 7000 rpm/s) for 25s. 70 uL of chlorobenzene (anti-solvent) was dropped gently on the substrate after 15s. After the substrates turned brown, they were placed on a pre-heated hotplate at 100 °C for 10 minutes. The thickness of the resulting films were ~ 200 nm.

## IV.  Device measurements:

FET measurements were performed using an Agilent 4155B parameter analyzer operated in pulsed mode (short impulse of 0.5 ms) for transfer characteristics and continuous mode for output characteristics. The temperature-dependent transport measurements were performed using a Desert Cryogenics low-temperature probe station**.**

## V.  Ab-initio calculations

We performed all *ab initio* calculations using density functional theory (DFT)-based method as implemented in the Vienna Ab initio Simulation Package (VASP 5.3.5).[57] For the plane-wave basis-set we consider a cut-off of 500 eV. The projected augmented wave (PAW) method was used to capture the ion-electron interactions.[58] We employed a generalized gradient approximations (GGA) in the form of Perdew-Burke-Ernzerhof functionals (PBE) to simulate the exchange-correlation interactions.[59] Note that, computation with PBE-GGA functional provide band gap values of halide perovskites that are close to the experimentally reported values. However, this agreement originates from the mutual cancelation of errors that appear due to the absence of spin-orbit coupling (bandgap narrowing) and electronic many-body interactions (bandgap widening) in these simulations.[46,47,60] We further include spin-orbit coupling (SOC) interactions as implemented in the VASP for calculating the band structures of all materials.[61] Note that, the inclusion of SOC significantly reduces the band gap of the halide perovskites. Initially, we optimize the cell volume and internal geometries of $FA_{0.875}Cs_{0.125}Pb_{(1-x)}Sn_xI_3$ without any constraints. For the ease of numerical calculation, we have used a A-cation composition as $FA_{0.875}Cs_{0.125}$ which is close to the experimentally optimized composition $FA_{0.85}Cs_{0.15}$. The atomic positions were then relaxed until the maximum force on each atomic sites is less than 0.01eV Å$^{-1}$. During these optimizations, we employ a 5×5×5 Γ-centered k-point mesh for Brillouin zone sampling. The DFT-D3 method as implemented by Grimme has also been considered to simulate the dispersion corrections.[62] For the electronic structure calculations, the Brillouin zone integrations were done in a Γ-centered7×7×7 k-point mesh with Gaussian smearing of 0.01 eV. To simulate the mixed B-metal lattices, we considered several orderings of Pb and Sn in a 2×2×2 lattice that has 8 B-sites. Then, we thoroughly investigate the $FA_{0.875}Cs_{0.125}Pb_{0.5}Sn_{0.5}I_3$ lattice for this. However, we do not find any specific ordering to be considerably stable in terms of total energy. This matches well with the previous simulation by Goyel et al.[42] Thus, to keep the computational cost of these simulations reasonable, we only considered the structure where alternative B-sites were filled by Pb or Sn in all three dimensions.

## VI. Material characterization

*Photoluminescence (PL)*: PL measurements were performed using a confocal microscope (WITec, Alpha RAS system). A fiber-coupled 405-nm continuous-wave laser (Coherent, CUBE) was focused onto the sample using 40× objective. The average power of the laser at its focal point was 0.5 µW. PL of the sample was collected in the reflection geometry from the same objective while the excitation laser beam from the reflection was blocked using a 415-nm long-pass filter.

*X-ray diffraction (XRD):* XRD measurements were performed in Bragg-Brentano geometry using a D8 Advance X-ray diffractometer (Bruker AXS, Karlsruhe, Germany) with Cu-Kα source (1.5418 Å). Data for the experiments were collected in a locked-coupled 1D-mode for 2θ between 5º and 35º, with a step size of 0.01032º. For temperature dependent measurements, diffraction measurement was performed for every 5K interval over 92 K to 302 K.

*Scanning electron microscopy (SEM):* The surface of the perovskite films was imaged using a Zeiss LEO 1550 FE-SEM apparatus, with a 2 kV acceleration voltage. It was ensured that the voltage levels for SEM measurement did not significantly impact the integrity of the perovskite thin films. Grain size analysis was performed using ImageJ software.

*X-ray photoelectron spectroscopy (XPS):* XPS measurements were performed by a monochromatic Al Kα X-ray source (hν = 1486.6 eV) using a SPECS PHOIBOS 150 electron energy analyzers with a total energy resolution of 500 meV. Conductive silver paint was used to connect the sample surface to the holder to avoid charge accumulation during XPS measurements. All the films were deposited on Si substrates in the same way as used for FET devices.

*Photothermal deflection spectroscopy (PDS)*: PDS measurement and analysis was performed in the similar way as used in the reference [29].